\documentclass[english,aps,prl,reprint,showpacs]{revtex4-1}
\usepackage{graphicx}
\usepackage{amssymb}
\usepackage{color}

\begin{document}


\title{An island of stability in a sea of fingers: emergent large-scale features of the viscous flow instability}


\author{Irmgard Bischofberger} 
\email{ibischofberger@uchicago.edu}
\affiliation{The James Franck and Enrico Fermi Institutes and The Department of Physics, The University of Chicago, Chicago, Illinois 60637, USA}

\author{Radha Ramachandran}
\affiliation{The James Franck and Enrico Fermi Institutes and The Department of Physics, The University of Chicago, Chicago, Illinois 60637, USA}

\author{Sidney R.  Nagel}
\affiliation{The James Franck and Enrico Fermi Institutes and The Department of Physics, The University of Chicago, Chicago, Illinois 60637, USA}


\date{\today}

\begin{abstract}
The displacement of a more viscous fluid by a less viscous one in a quasi-two dimensional geometry leads to the formation of complex fingering patterns. This fingering has been characterized by a most unstable wavelength, $\lambda_c$, which depends on the viscosity difference between the two immiscible fluids and sets the characteristic width of the fingers. How the finger length grows after the instability occurs is an equally important, but previously overlooked, aspect that characterizes the global features of the patterns. As the lower viscosity fluid is injected, we show that there is a stable inner region where the outer fluid is completely displaced. The ratio of the finger length to the radius of this stable region depends only on the viscosity ratio of the fluids and is decoupled from $\lambda_c$.

\end{abstract}


\maketitle



One of the earliest and most important examples of pattern formation is the viscous fingering instability, which occurs when a less viscous fluid displaces a more viscous one confined within a thin gap. Since the work of Saffman and Taylor in 1958 \cite{saffman1958penetration}, a large literature has been dedicated to interpreting the formation of these structures \cite{homsy1987viscous, paterson1981radial, chen1987radial, maaloy1985viscous, bensimon1986viscous, miranda1998radial, moore2002fluctuations, praud2005fractal, goyal2006miscible, al2012control, Cueto2014}. Understanding their evolution is important not only for the science of pattern formation \cite{couder2002viscous, daccord1986radial, casademunt2004viscous}, but also for industrial applications ranging from hydrology and petroleum extraction to sugar refining and carbon sequestration \cite{hill1952channeling, orr1984use, white2012co2, wang2004unstable}.

The majority of the work has been focused on understanding the onset of the instability, when a circular interface develops small finger-like protrusions \cite{tabeling1987experimental, lajeunesse19973d, miranda1998radial, moore2002fluctuations, goyal2006miscible}. These studies are typically done in a Hele-Shaw cell, which consists of two parallel plates separated by a thin gap, where a high viscosity fluid, $\eta_{out}$, is displaced by a lower viscosity fluid, $\eta_{in}$. Saffman and Taylor showed that there is a most unstable wavelength, $\lambda_c$, that depends on the difference in viscosity, $\Delta \eta \equiv \eta_{out} - \eta_{in}$, the interfacial tension, $\sigma$, the interfacial velocity, $V$, and the plate separation, $b$ \cite{saffman1958penetration}:
\begin{equation}
\lambda_c = \pi b\sqrt{\frac{\sigma}{\Delta \eta V}}.
\label{saff-t}
\end{equation}

\begin{figure}
\centering 
\begin{center} 
\includegraphics[width =\columnwidth]{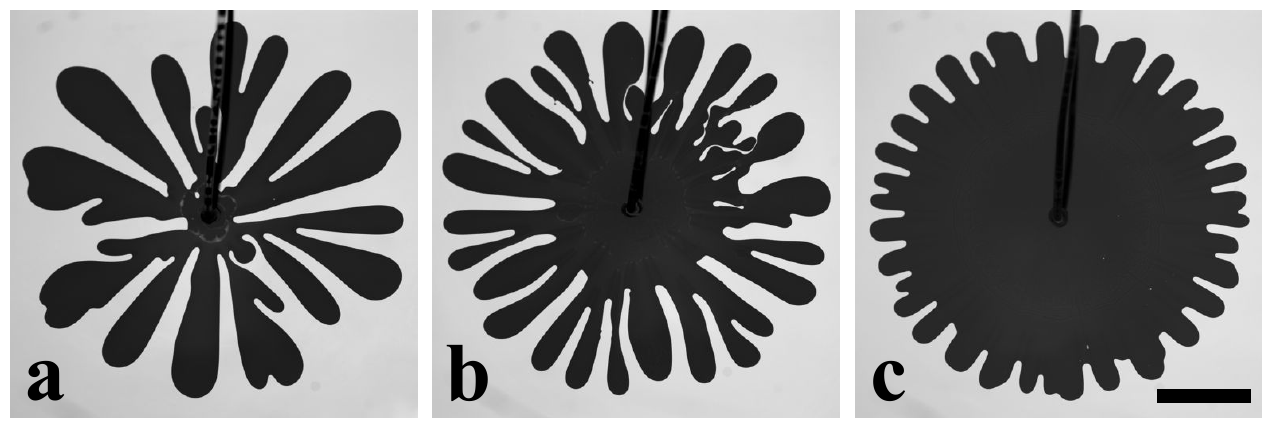} 
\caption{Fingering patterns showing the presence of an inner circular region of complete displacement of the outer fluid. The inner circle increases with increasing viscosity ratio of the inner to the outer fluid,  $\eta_{in}/\eta_{out}$, while the most unstable wavelength, $\lambda_c$, is approximately held constant. (a) $\eta_{in}/\eta_{out}$ = 0.0033 ($\eta_{out}$ = 296.7 mPa s, $\eta_{in} $ = 0.99 mPa s), (b) $\eta_{in}/\eta_{out}$ = 0.14 ($\eta_{out}$ = 345.1 mPa s, $\eta_{in}$ = 46.6 mPa s) (c) $\eta_{in}/\eta_{out}$ = 0.42 ($\eta_{out}$ = 530.3 mPa s, $\eta_{in}$ = 224.4 mPa s). In all cases, the colorless outer fluid is a silicone oil and the dyed inner fluid is a glycerol-water mixture. In these experiments, $\Delta \eta$ = 300$\pm$6 mPa s, $b$ = 254 $\mu$m, $q$ = 10 ml/min and $\sigma$ = 26.5$\pm$2.5 mN/m. The scale bar is 4 cm. }
\label{4lambda}
\end{center}
\end{figure}

However, these studies neglect the global features of the fingering patterns that are formed at late times after the fingers are fully developed. We find that these patterns are characterized not only by the finger width, but also by their length. In particular, we find an interior region in which the outer fluid is completely displaced. This provides a scale with which to measure the finger length as the growth proceeds. This is shown by the three images in Fig. \ref{4lambda}. In all cases, the viscosity difference, $\Delta \eta$, the interfacial tension, $\sigma$, the plate spacing, $b$, and the flow rate, $q$, are held nearly constant as given in the figure caption. The distinctively different patterns are created simply by varying the viscosity ratio, $\eta_{in}/\eta_{out}$. With increasing $\eta_{in}/\eta_{out}$, the inner circular region increases dramatically while the length of the fingers compared to the inner radius decreases.

Such behavior was previously noted in the singular case of miscible fluids where the interfacial tension is nearly zero \cite{irmgard}. In this paper, we show that this is a general result even for immiscible fluids. Moreover, we show that the inner region is independent of all the parameters that control the most unstable wavelength. This could not be done in experiments with miscible fluids where $\lambda_c$ in Eq. \ref{saff-t} is cut-off by the plate spacing, $b$ \cite{paterson1985fingering, lajeunesse19973d}.

The experimental set-up consists of two glass plates of 14 cm radius and 1.9 cm thickness to form the Hele-Shaw cell. The plates are maintained at a constant gap, $b$, by placing spacers of various thickness, ranging from 177 $\mu$m to 635 $\mu$m, at the edges of the plates. The liquids are injected through a 1.6 mm hole in the center of the plate using a syringe pump (New Era Pump Systems NE-1010) that maintains the volumetric flow rate at a set constant value. The patterns are recorded at frame rates ranging from 2 fps to 14 fps. 

In our experiments we use silicone oils (Clearco) of viscosities ranging between 98 mPa s and 1025 mPa s as the outer fluid. Both glycerol-water mixtures (Fisher Scientific) and mineral oils (Fisher Scientific) are used as the inner fluids. These two combinations have very different interfacial tensions: 24 mN/m - 29 mN/m for pairs of silicone oil/glycerol-water mixtures and 1 mN/m - 1.2 mN/m for pairs of silicone oil/mineral oil combinations. The inner fluids are dyed using Brilliant Blue G dye (Alfa Aesar) or Oil Red O dye (Sigma-Alrich) to enhance the contrast between the inner and outer fluids. Interfacial tensions, measured using the pendant drop method, are in agreement with literature values \cite{than1988measurement}.

\begin{figure}
\centering 
\begin{center} 
\includegraphics[width = 0.5\columnwidth]{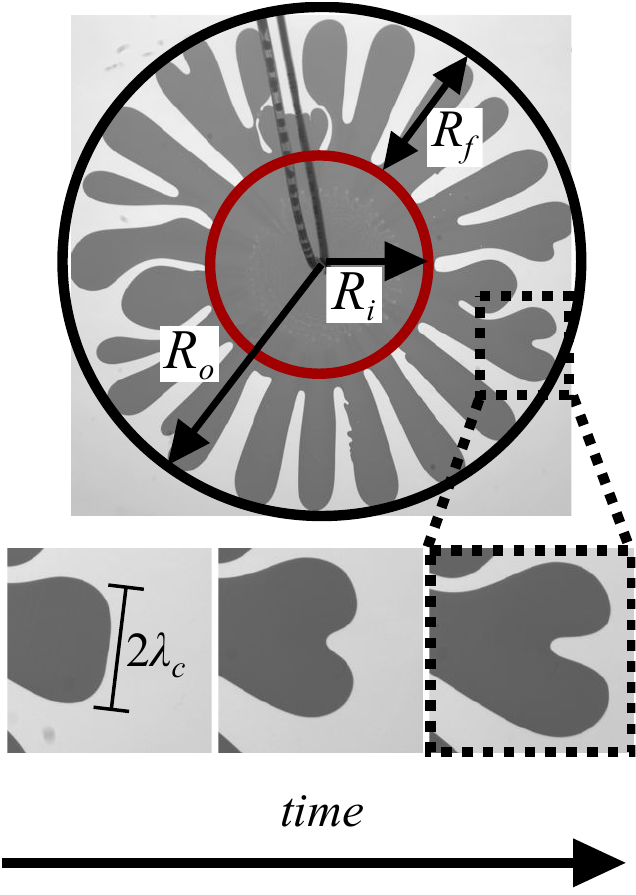} 
\caption{Determination of the characteristic length scales defining the large-scale structure and the most unstable wavelength, $\lambda_c$. The inner radius, $R_{i}$, is the radius of the largest circle completely inscribed in the inner fluid. The outer radius, $R_{o}$, is the radius of the smallest circle that encloses the entire pattern. The finger length is $R_{f} \equiv R_{o} - R_{i}$. $\lambda_c$ is defined as half the width of a finger just before it splits, as indicated in the time series of three images in the bottom panel. } 
\label{lambda-sizeratio}
\end{center}
\end{figure}

To characterize the overall shape of the patterns, we define three characteristic length scales as shown in the top panel of Fig. \ref{lambda-sizeratio}. We define an inner radius, $R_{i}$, which is the radius of the largest circle that fits completely inside the inner fluid and an outer radius, $R_{o}$, which is the smallest circle that encloses the entire pattern. The length of the finger is $R_{f}  \equiv R_{o} - R_{i}$. The size-ratio $\equiv R_{f}/R_{i}$. 

The most unstable wavelength, $\lambda_c$, is determined by half the width of a finger before a splitting event, as shown in the bottom panels of Fig. \ref{lambda-sizeratio}. To account for the decrease in velocity with the distance from the center of our radial cell, we only consider splitting events that occur within a radius of 3 cm to 5 cm from the nozzle.

\begin{figure}
\centering 
\begin{center} 
\includegraphics[width = 0.5\columnwidth]{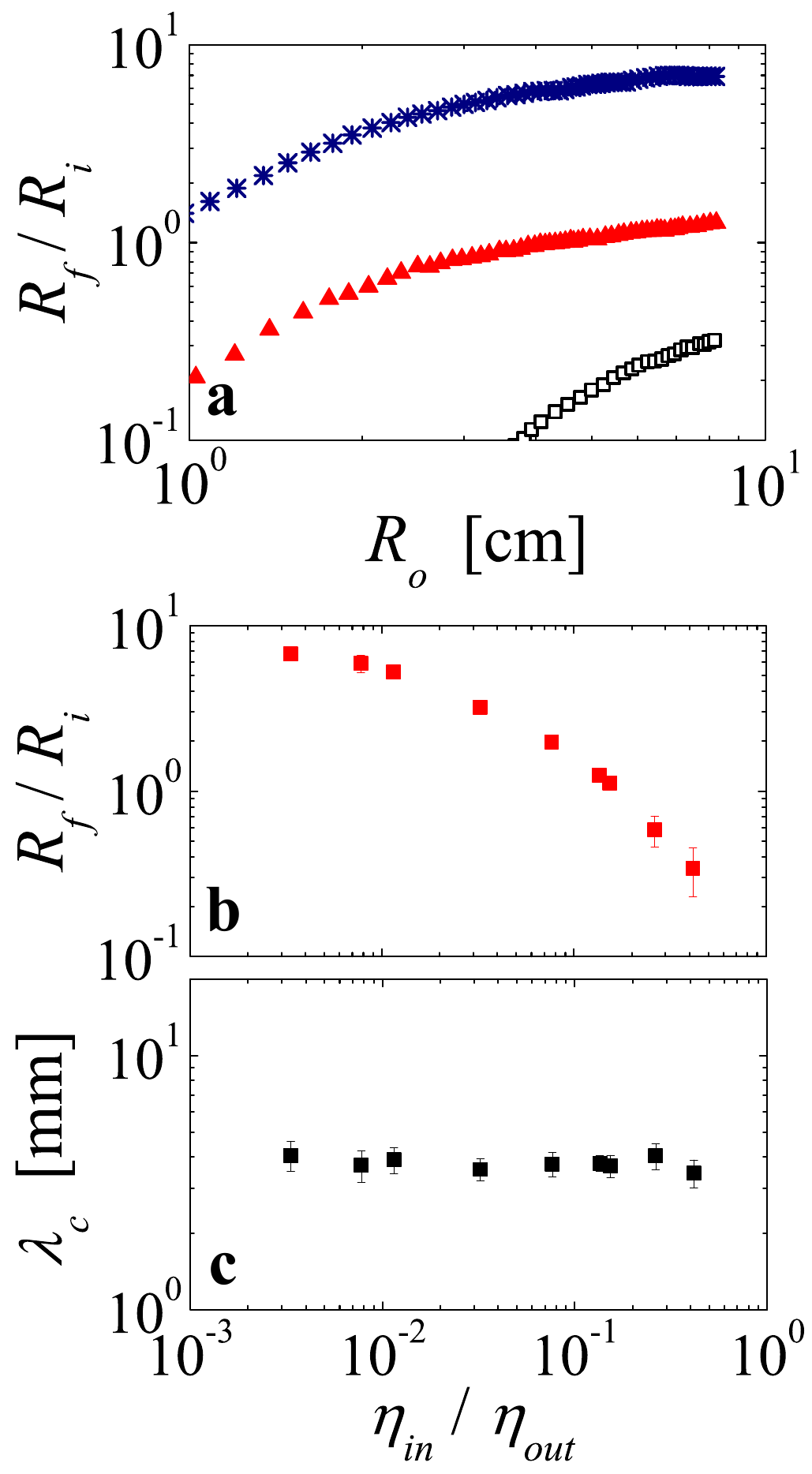} 
\caption{(a) Growth of the size-ratio, $R_f/R_i$, with the outer radius, $R_o$, for three different viscosity ratios: ($\ast$) $\eta_{in}/\eta_{out}$ = 0.0033 ($\eta_{out}$ = 296.7 mPa s, $\eta_{in} $= 0.99 mPa s), ($\blacktriangle$) $\eta_{in}/\eta_{out}$ = 0.14 ($\eta_{out}$ = 345.1 mPa s, $\eta_{in}$ = 46.6 mPa s), ($\Box$) $\eta_{in}/\eta_{out}$ = 0.42 ($\eta_{out}$ = 530.3 mPa s, $\eta_{in}$ = 224.4 mPa s). (b) and (c) The dependence of the size-ratio, $R_f/R_i$, and the most unstable wavelength, $\lambda_c$, on $\eta_{in}/\eta_{out}$. The size-ratio is measured at $R_o =$ 8 cm. In these experiments, $\Delta \eta$ is approximately constant  ($\Delta \eta =$ 300$\pm$6 mPa s) and $b$ = 254 $\mu$m, $q$ = 10 ml/min,  $\sigma$ = 26.5$\pm$2.5 mN/m. $R_f/R_i$ changes dramatically with viscosity ratio, while $\lambda_c$ is unchanged. }
\label{plot-samelambda}
\end{center}
\end{figure}

\begin{figure*}
\centering 
\begin{center} 
\includegraphics[width =1.8\columnwidth]{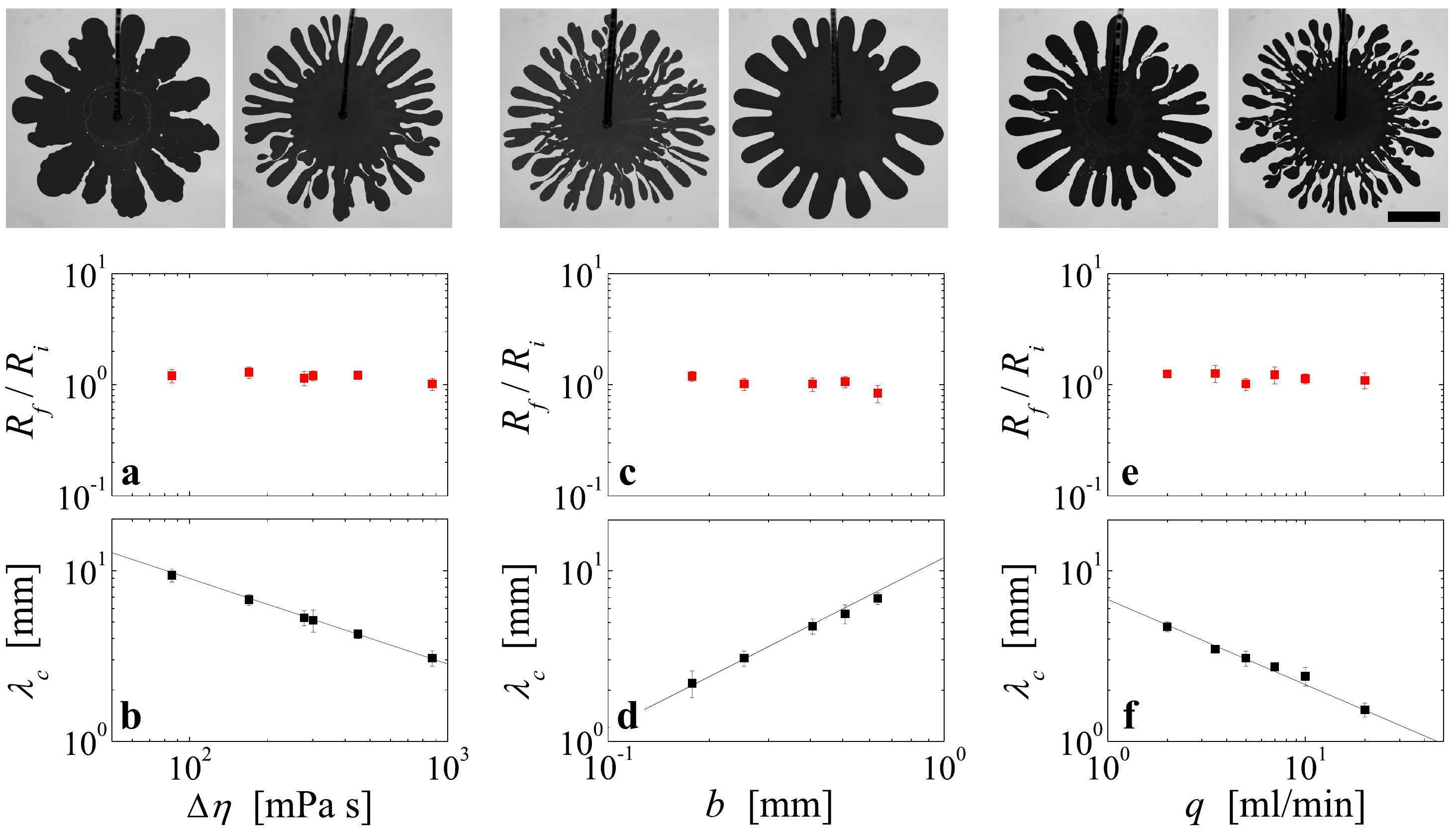} 
\caption{Dependence of the size-ratio, $R_f/R_i$, and the most unstable wavelength, $\lambda_c$, on viscosity difference, $\Delta \eta$, plate spacing, $b$, and flow rate, $q$. The three control parameters are changed one at a time, while keeping the others constant. (a) and (b): $\eta_{in}/\eta_{out}$ = 0.15; $\Delta \eta$ is varied while keeping $q$ = 5 ml/min,  $b$ = 254 $\mu$m and $\sigma$ = 26.5$\pm$2.5 mN/m. (c) and (d): $\eta_{in}/\eta_{out}$ = 0.15; $b$ is varied while keeping $\Delta \eta$ = 871.8 mPa s, $q/b$ = 196.9 sq.cm/min, and $\sigma$ = 26.5$\pm$2.5 mN/m. (e) and (f): $\eta_{in}/\eta_{out} = $ 0.15; $q$ is varied while keeping $\Delta \eta$ = 871.8 mPa s, $b$ = 254 $\mu$m, and $\sigma$ = 26.5$\pm$2.5 mN/m. The lines in (b, d and f) denote the predicted wavelengths according to Eq. \ref{saff-t}. In all cases, $R_f/R_i$ is unaffected while $\lambda_c$ varies in accord with Eq. \ref{saff-t}. The size-ratio is measured at $R_o =$ is 8 cm. The scale bar is 4 cm.} 
\label{allparams}
\end{center}
\end{figure*}

The growth of the patterns is characterized by an initial fast growth of $R_f/R_i$ that gradually slows down as the pattern grows larger, as shown in Fig. \ref{plot-samelambda}a for three different viscosity ratios (see supplementary movies). While the overall dependence is similar for all three data sets, the absolute value of $R_f/R_i$ is very different, indicating that the circular region grows by different amounts. Note that the radial geometry is known to produce a small delay in the onset of the instability which also leads to the appearance of a circular stable region. However, this effect is small for our experiments; this onset radius corresponds to approximately 1.5 mm (comparable to the central injection hole in our plates) \cite{al2012control, miranda1998radial}. In particular, we here focus on the temporal evolution of the inner radius, $R_i$, which continues long after the onset of fingering.

In order to compare the patterns formed at different viscosity ratios, we measure $R_f/R_i$ when the outer radius $R_o$ reaches 8 cm. In Fig. \ref{plot-samelambda}b we show that $R_f/R_i$ measured at this value decreases rapidly with increasing $\eta_{in}/\eta_{out}$. As $\eta_{in}/\eta_{out}$ approaches 1, the system goes towards a completely stable displacement. In the other limiting case, when $\eta_{in}/\eta_{out}$ approaches 0, the patterns have a vanishingly small inner stable region. In all of this data, we have varied $\eta_{in}/\eta_{out}$ in such a way as to leave $\Delta \eta$ nearly constant ($\Delta \eta =$ 300$\pm$6 mPa s). We have also kept $q$, $b$ and $\sigma$ the same. As shown in Fig. \ref{plot-samelambda}c, the most unstable wavelength does not change under these conditions, whereas $R_f/R_i$ drops by more than a decade and a half.

\begin{figure}
\centering 
\begin{center} 
\includegraphics[width = 0.85\columnwidth]{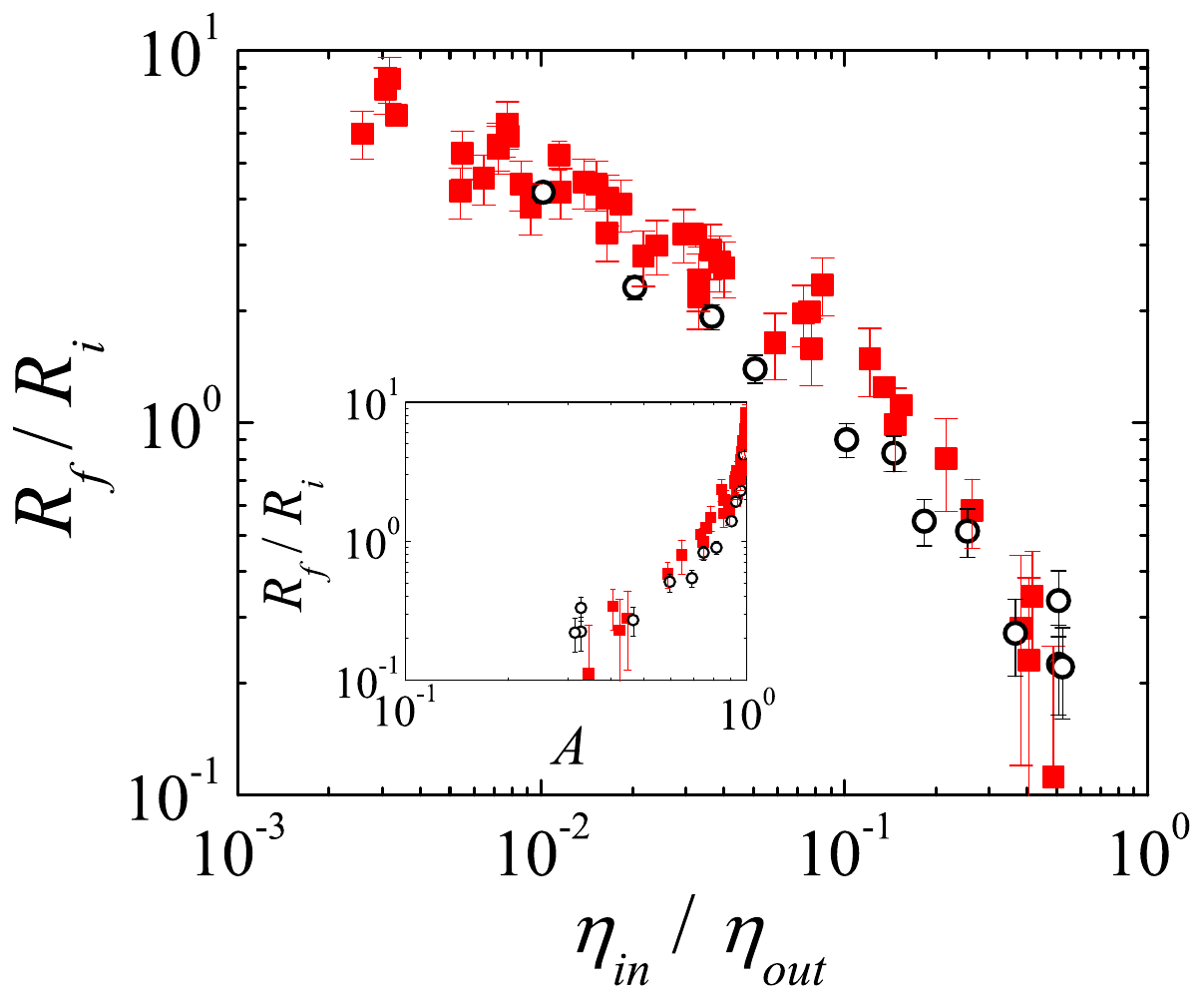} 
\caption{Dependence of the size-ratio, $R_f/R_i$, on viscosity ratio, $\eta_{in}/\eta_{out}$, for two sets of fluids with different interfacial tensions, $\sigma$. The squares are for pairs of silicone oil/glycerol-water mixture with $\sigma$ = 24 mN/m - 29 mN/m, the open circles are for pairs of silicone oil/mineral oil with $\sigma$ = 1 mN/m - 1.2 mN/m. The size-ratio is measured at $R_o = 8$ cm. The two data sets are nearly the same, even though the interfacial tensions differ by a factor of approximately 25. Inset: the same $R_f/R_i$ data plotted versus $A \equiv (\eta_{out}-\eta_{in})/(\eta_{out}+\eta_{in})$.} 
\label{2wavelength}
\end{center}
\end{figure}

$R_f/R_i$ clearly depends on the viscosity ratio, $\eta_{in}/\eta_{out}$. To understand what other parameters it might depend on, we systematically vary one of the control parameters at a time, while leaving the others fixed. The results are shown in Fig. \ref{allparams}. In the left-most column, we show that a change in $\Delta \eta$ by a decade leaves $R_f/R_i$ unchanged. The most unstable wavelength, however, decreases with increasing $\Delta \eta$ consistent with Eq. \ref{saff-t}. In the middle and right columns, we show similar behavior when we vary the plate spacing, $b$, and the flow rate, $q$. In all cases, $R_f/R_i$ remains unchanged while the most unstable wavelength continues to vary in agreement with the Saffman-Taylor prediction, as indicated by the solid lines in Fig. \ref{allparams}b,d,f which correspond to Eq. \ref{saff-t}. The agreement between Eq. \ref{saff-t} and our experimental data further implies that the wavelength selection at tip-splitting events in the non-linear regime follows the same dependences as that governing the instability onset in the linear regime. The images at the top of each column show the patterns at the two extreme values of $\Delta \eta$, $b$ and $q$; it is strikingly obvious that they have nearly constant inner radii but very different finger widths.

Figure \ref{2wavelength} shows the dependence of the size-ratio on viscosity ratio for two sets of fluids, glycerol-water mixtures/silicone oils and mineral oils/silicone oils. These fluids have interfacial tensions that differ by approximately a factor of 25. The two data sets are essentially indistinguishable, indicating that $R_f/R_i$ is independent of $\sigma$ and only controlled by $\eta_{in}/\eta_{out}$. The two liquids used to displace the silicone oils have different wetting properties; mineral oil fully wets the glass while glycerol-water mixtures only partially wet glass. That the patterns formed by the two sets of fluids exhibit the same dependence on $\eta_{in}/\eta_{out}$ indicates that the wetting properties do not affect the pattern growth. Lastly, we note that $R_f/R_i$ is further independent of the size of the Hele-Shaw cell; patterns formed in a cell of 25 cm radius are indistinguishable from those formed  in a cell of 14 cm radius.

The importance of the relative values of the inner and outer viscosities is in principle also captured in a different commonly used combination of the viscosities: $A \equiv (\eta_{out}-\eta_{in})/(\eta_{out}+\eta_{in})$ \cite{petitjeans1996miscible, chen1996miscible}. Clearly, this is just a function of $\eta_{in}/\eta_{out}$. The question is, which is the more natural variable to describe the behavior of the size ratio. In the inset of Fig. \ref{2wavelength}, we plot the same data for $R_f/R_i$ against $A$. As $A$ approaches 1, the size-ratio increases rapidly within a very small range of $A$. This suggests that $A$ is a less-suitable parameter than $\eta_{in}/\eta_{out}$ to describe the observed global features of the instability.

Taken together, the data in Figs. 3, 4 and 5 show that while the most unstable wavelength is indeed set by Eq. \ref{saff-t}, there is a second parameter that characterizes the large-scale features of the patterns. This second parameter, $R_f/R_i$, is completely independent of the most unstable wavelength, $\lambda_c$. It is only set by the viscosity ratio, $\eta_{in}/\eta_{out}$.

Looking at the images presented in this paper, three features immediately pop out and grab our attention. First is the presence of fingers that have characteristic widths given by the most unstable wavelength, $\lambda_c$. This feature has been studied thoroughly both experimentally and theoretically. Equally apparent is the presence of a large circular region where the outer fluid is completely displaced. Finally, we see a large variation in the length of the fingers. Our data show that the finger length is unrelated to the finger width and is set by only two parameters, the viscosity ratio, $\eta_{in}/\eta_{out}$, and the size of the inner circular region, $R_i$. All these experiments have been conducted in a radial Hele-Shaw cell. It is important to see if the large-scale features found here are also present in a linear geometry.

Such an interior region of complete displacement was previously found for miscible fluids where the interfacial tension is negligible \cite{irmgard}. This study raised the question whether this feature resulted from the peculiarity of the singular zero-interfacial-tension limit \cite{cheng2008towards, nittman1985fractal, lee2006bubble}. The experiments presented here show that this result is very general, applying to both miscible and immiscible pairs of fluids. In particular, with the immiscible fluids studied here, we are able to vary $\lambda_c$ and show conclusively that these two important features of the pattern are independent of one another.

\begin{acknowledgments}
We thank Rudro Rana Biswas, Justin Burton, Todd Dupont, Julian Freed-Brown, Leo Kadanoff, Paul Wiegmann, Tom Witten and Wendy Zhang. This work was supported by NSF grant DMR-1404841. I.B. gratefully acknowledges financial support from the Swiss National Science Foundation (PBFRP2-134287).
\end{acknowledgments}

\bibliographystyle{apsrev4-1}
\bibliography{island}

\end{document}